\documentclass[twocolumn,showpacs,nopreprintnumbers,amsmath,pra,superscriptaddress]{revtex4}
\usepackage{amsmath,amssymb}

\usepackage{graphicx}

\begin{document}

\title{On the unique possibility to increase significantly the contrast of dark resonances on D1 line of $^{87}$Rb.}

\author{A.V. Taichenachev}
\affiliation{Novosibirsk State University, Novosibirsk 630090, Russia} \affiliation{Institute of Laser Physics SB
RAS, Novosibirsk 630090, Russia}
\author{V.I. Yudin}
\affiliation{Novosibirsk State University, Novosibirsk 630090, Russia} \affiliation{Institute of Laser Physics SB
RAS, Novosibirsk 630090, Russia}
\author{V.L. Velichansky}
\affiliation{P.N.Lebedev Physical Institute RAS, Moscow 117924, Russia}
\author{S.A. Zibrov}
\affiliation{P.N.Lebedev Physical Institute RAS, Moscow 117924, Russia}

\begin{abstract}
We propose and study, theoretically and experimentally, a new scheme of excitation of a coherent
population trapping resonance for D1 line of alakli atoms with nuclear spin $I=3/2$
by bichromatic linearly polarized light ({\em lin}$||${\em
lin} field) at the conditions of spectral resolution of the excited state.
The unique properties of this scheme result in a high contrast of dark resonance for D1 line of $^{87}$Rb.
\end{abstract}


\maketitle

Great attention is drawn to the study of the coherent interaction of atoms with electromagnetic fields. Under
specific experimental conditions coherent atom-field interaction result in what is called the coherent population
trapping effect (CPT) \cite{ar}. Narrow-width CPT resonances ("dark resonances") induced by this effect are used
in atomic clocks \cite{Hem,Kit,Li}, precise magnetometers \cite{St1,Sch}, laser cooling, spectroscopy, optical
signal processing, etc. Most applications require CPT resonance with optimized parameters such as large
amplitude, small spectral width and small background simultaneously. In addition minimized light shifts are
needed in metrology.

Most of theoretical and experimental studies of the CPT phenomena are made for alkali atoms. The CPT resonance is
usually observed with a bichromatic field consisting of two resonant laser fields with the frequency difference
changing in the vicinity of the hyperfine splitting of the ground state. At exact two-photon resonance atoms are
optically pumped into a coherent nonabsorbing superposition of the ground states which is refereed to as a dark
state. A variety of different excitation schemes of the CPT resonances were proposed. They differ by the choice
of the isotope, excitation line (D1 or D2), the field characteristics (detuning, amplitude, and polarization of
the field components). Different techniques increasing coherent atom-field interaction time also play an important
role on the build up of the ground state coherence. These techniques include the use of a mixture of an alkali
vapor with different buffer gases or/and use of cells with antirelaxation coating of the inner walls.

Experiments \cite{Br,Win} carried out with the cell containing a mixture of Cs and a buffer gas revealed a
resonance with a linewidth as small as 50 Hz in the case of D2 line excitation. However, the contrast of the CPT
resonance in these experiments was only a small  fraction of a percent. In \cite{St} it was shown that using the
D1 line instead of D2 line for excitation significantly increases the contrast of the dark resonance. This fact
was later confirmed by other authors \cite{Van}. In these experiments resonances were produced by circularly
polarized light fields ($\sigma$-$\sigma$) which induced two-photon transitions between magnetic sublevels with
the same quantum number m (m--m resonances). In atomic clocks the 0--0 transition is used because it is first order
insensitive to the magnetic field. The D1 line provides  better resonance than the D2 line since at exact
two-photon resonance in D1 line a dark superposition state exists even when hyperfine components of the excited
state are not spectrally resolved because of buffer gas collisional broadening (on the contrary, the D2 line in
such a case destroys dark state via cycling transitions and the dark state is
absent in such a case). When the atom-field interaction operator is applied to this state it gives zero:
\begin{equation}\label{dark}
-(\widehat{\bf d}{\bf E})|dark\rangle=0
\end{equation}
This $|dark\rangle$ state is a coherent superposition of wave functions of the Zeeman sublevels of the ground
state. Atoms in this state neither absorb nor emit light.

However, the D1 line has its own limitations. In the case of $\sigma$-$\sigma$ laser fields a state insensitive to
two-photon detuning $|trap\rangle$ exists (in the extreme Zeeman sublevel). A fraction of atoms accumulates
in this $|trap\rangle$ state due to optical pumping and does not contribute to the ground state coherence. This
fact significantly limits the contrast of the resonance. Thus, to optimize parameters of the dark resonance one
has to find such excitation schemes that on the one hand have a $|dark\rangle$ state and on the other hand do not
have a $|trap\rangle$ state.

Recently this problem has been solved for the case of $m$--$m$ resonances on D1 line. Different approaches for
optimizing parameters of the dark resonance were proposed in \cite{Taich,Jau,Zan} for the case of 0--0 resonance.
A general case of arbitrary $m$--$m$ resonance is investigated in \cite{Taich1}. All these methods use resonant
field components with different polarizations. For example, in \cite{Jau,Zan} the two components of bichromatic
radiation have orthogonal linear polarizations ({\em lin}$\perp${\em lin}). In \cite{Taich} spatially dependent
polarization of the light field can be considered as two-frequency {\em lin}$\perp${\em lin} configuration, which
is formed locally in this case as compared to \cite{Jau,Zan}.

In this paper we propose an alternative method of forming high-contrast dark resonances by a bichromatic light
field with arbitrary polarizations (in general case - elliptical). This method works only in a special case when
total angular moments of the ground states are $F_g$=1,2 and the coupling occurs through an excited state with a
total angular momentum of $F_e$=1. Besides, a key requirement here is a good spectral resolution of the excited
hyperfine levels. Such a situation is realized for the D1 line of 87Rb because hyperfine splitting of the excited
state in this case is 812 MHz and is considerably larger than the Doppler width of the line ($\sim$400 MHz). A CPT
resonance involves in this case two pairs of Zeeman sublevels who's magnetic quantum number differ by two:
($-$1)--($+$1) and ($+$1)--($-$1). At exact two-photon resonance the dark superposition states exist, while an
extra trapping state $|trap\rangle$ is absent. The same situation occurs with field-sensitive resonances
($+1$)--($+1$) and ($-1$)--($-1$) which also have high contrast in this case. Special attention is drawn to the
case of {\em lin}$||${\em lin} configuration of optical fields in which frequency components have the same linear
polarizations. It should be noted that due to the nuclear contribution of the $g$-factor of different
ground-state hyperfine components these resonances ($-$1)--($+$1) and ($+$1)--($-$1) have a weak linear
sensitivity to a magnetic field. However, for the {\em lin}$||${\em lin} light field configuration, owing to the
symmetry, this effect manifests mostly in a small broadening of the resonance lineshape. It is important that the
center of the resonance has zero linear sensitivity to a magnetic field. Qualitative theoretical analysis is
proved by experiments which demonstrate a very high contrast of the dark resonance in case of {\em lin}$||${\em
lin} excitation scheme. Also the influence of nuclear spin, which leads to the broadening and spliting of the
resonance, is studied.

{\bf Qualitative theoretical picture.} Let's consider the resonant interaction of atoms with bichromatic field,
consisting of two co-propagating (along axis $z$) running waves
\begin{equation}\label{1}
{\bf E}(z,t)={\bf E}_1(z)e^{-i\omega_1t}+ {\bf E}_2(z)e^{-i\omega_2t}+c.c.
\end{equation}
The two components of this field have arbitrary amplitudes and polarizations (elliptical in general case). The
polarization can be presented as a linear combination of two orthogonal circular components (${\bf e}_{\pm
1}=\mp({\bf e}_x\pm i{\bf e}_y)/\sqrt{2}$) in the following way:
\begin{equation}\label{2}
{\bf E}_j=E^{(j)}_{-1}{\bf e}^{}_{-1}+E^{(j)}_{+1}{\bf e}^{}_{+1};\quad (j=1,2).
\end{equation}
A small static longitudinal magnetic field {\bf B} is applied along the $z$ axis of the coordinate system. When
the frequency difference ($\omega_1-\omega_2$) of the bichromatic field components is equal to the hyperfine
splitting of the ground state $\Delta_{hfs}$, this field induces a two-photon dark resonance of a $\Lambda$-type
on two hyperfine levels of the ground state.

Let's take in account only the alkali atoms with a nuclear spin number $I$=3/2 ($^{7}$Li, $^{23}$Na, $^{39,41}$K
and $^{87}$Rb). The total angular momenta of the hyperfine components of the ground state for these atoms are
$F_g$=1,2. We consider the case when both frequency components of the bichromatic field are resonant with the
common excited state with a total angular momentum $F_e$=1. The excitation scheme for this case is shown in figure
Fig.1{\em a}.

Fig.1{\em a} shows that a two-photon resonance of a $\Lambda$-type is formed in two pairs of ground-state
hyperfine sublevels with $|F_g$=1,{\em m}=$-1\rangle$, $|F_g$=2,{\em m}=$+1\rangle$ and $|F_g$=1,{\em
m}=$+1\rangle$, $|F_g$=2,{\em m}=$-1\rangle$. Further in the paper we will call them as ($-$1)--($+$1) and
($+$1)--($-$1) resonances. Both of these $\Lambda$-schemes are excited through a common excited state
$|F_e$=1,{\em m}=$0\rangle$. It is well known that for alkali atoms $g$-factors of different hyperfine components
of the ground state have an equal absolute value and opposite signs. In our case $g$-factors are $g$=$\pm$1/2. If
the contribution of the nuclear spin to the Zeeman splitting of levels is neglected, then two-photon resonances
(both ($-$1)--($+$1) and ($+$1)--($-$1)) frequencies are equal to the frequency of the 0--0 resonance formed on
$|F_g$=1,{\em m}=$0\rangle$ and $|F_g$=2,{\em m}=$0\rangle$ states. This fact means that at least two
superposition dark states exist in case of exact two-photon resonance for fields with arbitrary polarizations.
These states are determined by the above mentioned  $\Lambda$-schemes:
\begin{eqnarray}\label{D_pm}
&&|dark(\pm)\rangle=\\
&&N_{\pm}\left\{|F_g=1,m=\mp 1\rangle -\frac{V^{(1)}_{\pm 1}E^{(1)}_{\pm 1}}{V^{(2)}_{\mp 1}E^{(2)}_{\mp 1}
}\,|F_g=2,m=\pm 1\rangle\right\} .\nonumber
\end{eqnarray}
Here $V^{(1,2)}_{\pm 1}$ are the corresponding matrix elements of the dipole moment operator (Fig.1{\em a}),
$N_{\pm}$ are the normalization constants. In this general case a $|trap\rangle$ state is absent. This fact gives
the ability of forming a high contrast resonance. It is worth mentioning that resonance's shift due to the second
order magnetic field dependence in case of this scheme of excitation is 1.33 times smaller than for the true 0--0
resonance in case of traditional excitation scheme.

States given by (\ref{D_pm}) do not appear in any other scheme of excitation of the dark resonance. For example,
Fig.1{\em b} illustrates the situation, when excitation occurs through an excited state with $F_e$=2. Here ground
state levels $|F_g$=1,{\em m}=$\pm 1\rangle$ and $|F_g$=2,{\em m}=$\pm 1\rangle$ are coupled with the excited
states $|F_e$=2,{\em m}=$\pm 2\rangle$. These transitions destroy the dark states. For atoms with other quantum
numbers of total angular momentum $F_g$ (i.e. for those that have {\em I}$\neq$3/2) dark states (\ref{D_pm}) are
not induced for any given $F_e$.

Thus, alkali atoms with the nuclear spin of {\em I}=3/2 are of great interest due to new possibilities to form
high contrast magneto-insensitive dark resonances with arbitrary polarized light fields. However, the key
requirement of good spectral resolution of the excited state $F_e$=1 determines the choice of an atom and
transition. The most promising is the D1 line of $^{87}$Rb because the hyperfine splitting of the excited state is
equal to 812 MHz. This fact allows its use in cells at room temperatures. For all other alkali atoms
($^{7}$Li, $^{23}$Na, $^{39,41}$K) with similar structure, and also for D2 line of $^{87}$Rb good spectral
resolution of the excited state can be reached only in the case of laser cooled atoms or in a collimated atomic
beam.

We focus our attention on the case of {\em lin}$||${\em lin} polarizations of the light fields, where both
resonant fields have equal linear polarizations. For this excitation scheme $|${\em trap}$\rangle$ state does not
exist. Also, it is experimentally easy to create such a linear polarized field using one laser source. Besides,
this situation is preferable if the influence of nuclear spin on the splitting of levels is taken in account.
Nuclear spin leads to a small difference in absolute values of $g$-factors for different hyperfine components of
the ground state. As a result of this mismatch of $g$-factors the frequencies of the ($-1$)--($+1$) and
($+1$)--($-1$) two-photon $\Lambda$-resonances are also different (for $^{87}$Rb this difference is 2.8 kHz/G),
but their position is symmetrical relatively to the hyperfine splitting frequency $\Delta_{hfs}$. Thus, in case
of {\em lin}$||${\em lin} excitation scheme in the presence of a small magnetic field this fact leads only to the
broadening of the resonance and not to its shift (in linear approximation), because both $\Lambda$-system are
identical for {\em lin}$||${\em lin} field. Also, it is worth mentioning that the true 0--0 resonance in case of
{\em lin}$||${\em lin} excitation is absent due to a destructive interference of two-photon transitions induced by
different components of the circularly polarized light.

Let us mention another interesting peculiarity of the excitation scheme through the $F_e$=1 excited state. As
could be seen from Fig.1{\em a} two field sensitive resonances ($-1$)--($-1$) and ($+1$)--($+1$), formed on the
levels with the same $m$-quantum number, are also excited in $\Lambda$-type schemes. For these $\Lambda$-schemes
dark states exist at exact two-photon resonance. It means that these resonances can also have large contrasts.
This situation takes place for D1 line not only in the case of nuclear spin {\em I}=3/2, but in a more general
case of $F_g$=$F$,$F+$1, when resonant interaction of the two-frequency field (\ref{1}) occurs through level with
$F_e$=$F$. As can be seen from Fig.1{\em c} interaction between $|F_g$={\em F,m}=$-F\rangle$ and $|F_g$={\em
F}+1,{\em m}=$-F\rangle$, and also between $|F_g$={\em F,m}=$F\rangle$ and $|F_g$={\em F}+1,{\em m}=$F\rangle$
occurs by $\Lambda$-type scheme of excitation with arbitrary polarizations. Thus, resonances sensitive to the magnetic field
($-F$)--($-F$) and ($F$)--($F$) can also have large contrast. For example, for $^{133}$Cs atoms ({\em
I}=7/2) they will be ($-3$)--($-3$) and ($+3$)--($+3$) resonances under excitation through the level with $F_e$=3.

The given qualitative description is confirmed by numerical calculations and experiment.

{\bf Experiment.} Fig.2 shows a schematic of experimental setup. It includes a laser system, providing bichromatic
radiation, a heated cell with alkali metal vapor, and a signal detection scheme.

The experiment was carried out with a Pyrex cylindrical cell (40 mm long and 25 mm in diameter) containing
isotopically enriched $^{87}$Rb and 4 Torr Ne buffer gas. The cell was placed inside a solenoid that provided a
longitudinal magnetic field of 150 mG to lift the degeneracy of the Zeeman sublevels and to separate
magneto-insensitive ``clock'' resonance, which has no first-order magnetic field dependence, from the
field-dependent resonances. The solenoid was placed within three concentric $\mu$-metal shields in order to reduce
the influence of the external magnetic fields. The cell was heated by bifilar wire wrapped around the inner layer
of the magnetic shielding. For the experiments reported here the cell temperature was 50$^{\circ}$C.

In order to obtain a bichromatic light field, consisting of two resonant optical fields, we modulated the
injection current of a diode laser. It was done in the following way: a slave diode laser, operating in the
vicinity of 795 nm, was injection locked by the radiation of single mode extended cavity diode laser (ECDL) and
at the same time its current was frequency modulated at 3.417 GHz. This technique produced phase correlated
narrow linewidth optical fields and allowed easy tuning of these fields to the desired transitions. Also, it
allowed varying the resonant field's intensity ratio by changing the current of the injection-locked diode laser.
For the experiments reported here approximately 25\% of the total laser power was transferred to each first order
sideband, with the remainder residing in the carrier and high order sidebands. The laser's injection current was
frequency swept.

In the case of {\em lin}$||${\em lin} excitation scheme CPT resonance was formed by two linearly polarized first
order sidebands tuned to the $F_g$=1$\to$$F_e$=1 and $F_g$=2$\to$$F_e$=1 transitions. In case of $\sigma$-$\sigma$
excitation scheme these fields were circularly polarized and tuned to $F_g$=1$\to$$F_e$=2 and $F_g$=2$\to$$F_e$=2
transitions. The laser beam passed a quarter wave plate situated in a way to leave the polarization of light
unchanged (linear) or in a way to make it circular depending on the excitation scheme being studied. Optical
power (containing all sidebands) transmitted through the gas cell then was detected by a silicon photodiode.

We have performed a comparative study of two different excitation schemes of CPT resonances. CPT resonances
observed by use of different excitation schemes under the same experimental conditions are shown in Fig.3. Lines
are vertically shifted for the ease of visual perception. It is clearly seen from this figure that amplitude of
the clock resonance in case of the {\em lin}$||${\em lin} scheme of excitation through $F_e$=1 is much larger than
in case of $\sigma$-$\sigma$ excitation through $F_e$=2 and {\em lin}$||${\em lin} excitation through $F_e$=2. The
amplitude, the amplitude to width ratio, the contrast of the resonance in case of different excitation schemes
are shown in Fig.4,5,6.

Fig.7 shows the behavior of the clock CPT resonance in case of {\em lin}$||${\em lin} excitation scheme through
$F_e$=1 at different values of applied longitudinal magnetic field. In case of small magnetic fields this leads
to the broadening of the resonance and does not change the position of its maximum (in linear approximation). At
large magnetic field the resonance is eventually split into two components, because of the difference in absolute
values of $g$-factors for the two hyperfine levels. The center of the resonance line shape is shifted due to
quadratic Zeeman effect. It should be noted that the double structure has been observed previously
\cite{knappe99,levi02}. However, the observed resonance amplitudes were very small, because different excitation
schemes were used: D2 line of Cs in \cite{knappe99} and D1 line of $^{87}$Rb at higher buffer gas pressure in
\cite{levi02}. Recently it was proposed to use the gap between two peaks to lock the frequency in atomic
clock\cite{kazakov05}.

Thus, in the present paper we have proposed and realized experimentally the new scheme of excitation of D1 line
of alkali atoms with the nuclear spin $I$=3/2 by the bichromatic {\em lin}$||${\em lin} field. This allows us to
significantly increase the contrast of the magneto-insensitive dark resonance in comparison with the standard
excitation of the 0--0 resonance by the bichromatic circular polarized field. In usual conditions such a
possibility exists only for $^{87}$Rb, when the exited-state hyperfine levels are spectrally resolved. Besides,
in this case the magneto-sensitive $m$--$m$ resonances with maximal $|m|$ have the highest contrast. The obtained
results argue that D1 line of $^{87}$Rb under the excitation through the level with $F_e$=1 is a special object
of study for atomic clock and magnetometer based on CPT. These results have been firstly presented in part on
ICONO'05 \cite{icono05}.

We thank L. Hollberg, H. Robinson, J. Kitching, S. Knappe, V. Shah, and F. Levi for helpful discussions. This work
was supported by RFBR (05-02-17086, 04-02-16488, 05-08-01389) and by grant INTAS-01-0855. VLV and SAZ were
supported by grant ISTC 2651p.

e-mail: llf@laser.nsc.ru

\newpage

\begin{figure*}[h]\centerline{\scalebox{0.75}{\includegraphics{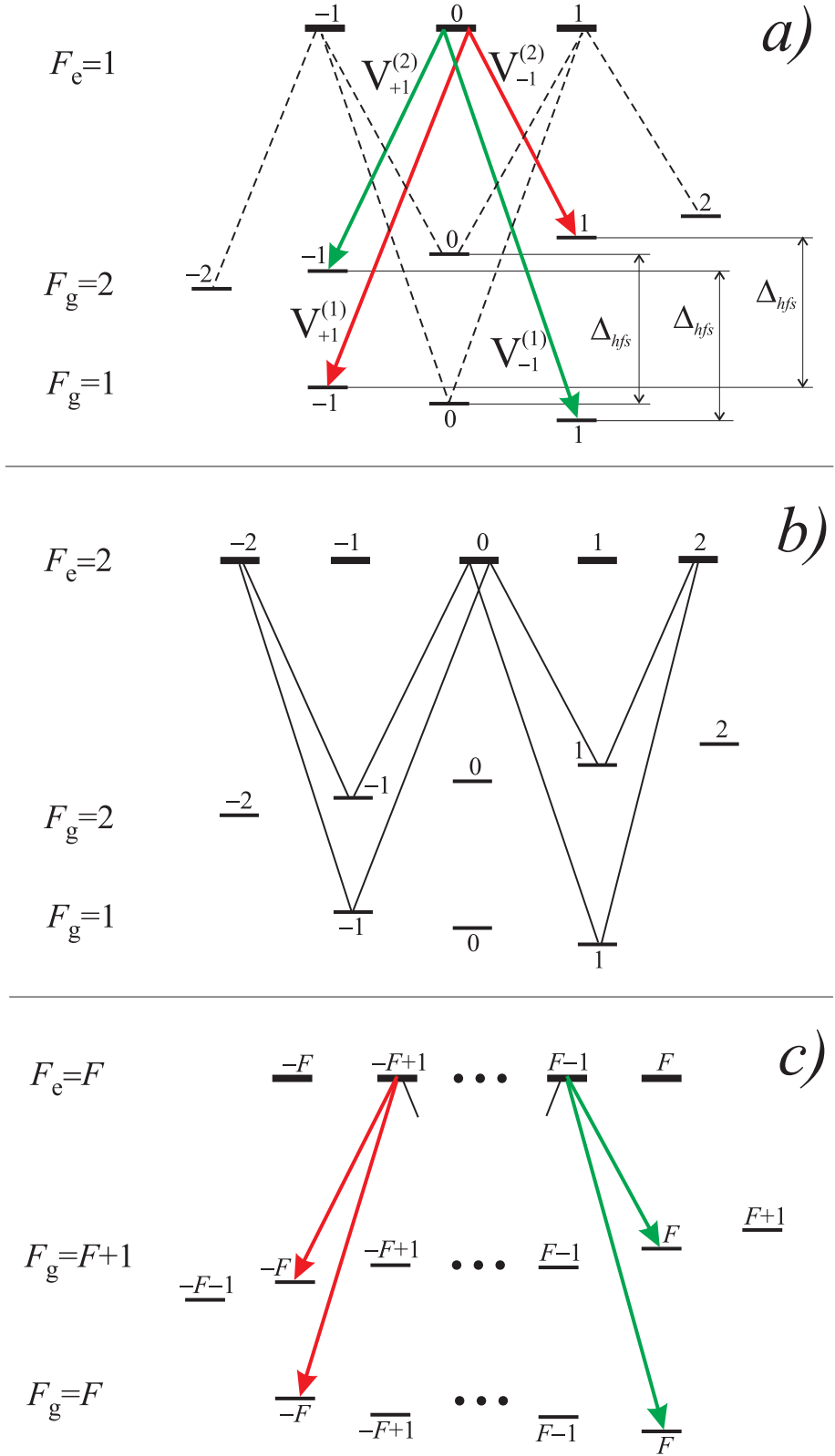}}}
\caption{Schemes of the light-induced transitions in a bichromatic field with arbitrary (elliptical in general)
polarizations for different variants of the excitation: {\em a)} The unique case of $F_g$=1,2 at the excitation
via $F_e$=1, when simple $\Lambda$-systems are realized for the Zeeman sublevels with different magnetic quantum
numbers $m$=$\pm$1 (solid color arrows); {\em b)} The case of $F_g$=1,2 at the excitation via $F_e$=2, when simple
$\Lambda$-systems are absent; {\em c)} The general case of $F_g$=$F$,$F$+1 at the excitation via $F_e$=$F$, when
the outermost magneto-sensitive resonances are formed through simple $\Lambda$-systems (color arrows) for
arbitrary (elliptical in general) polarizations.} \label{fig1}
\end{figure*}

\begin{figure*}[h]\centerline{\scalebox{0.75}{\includegraphics{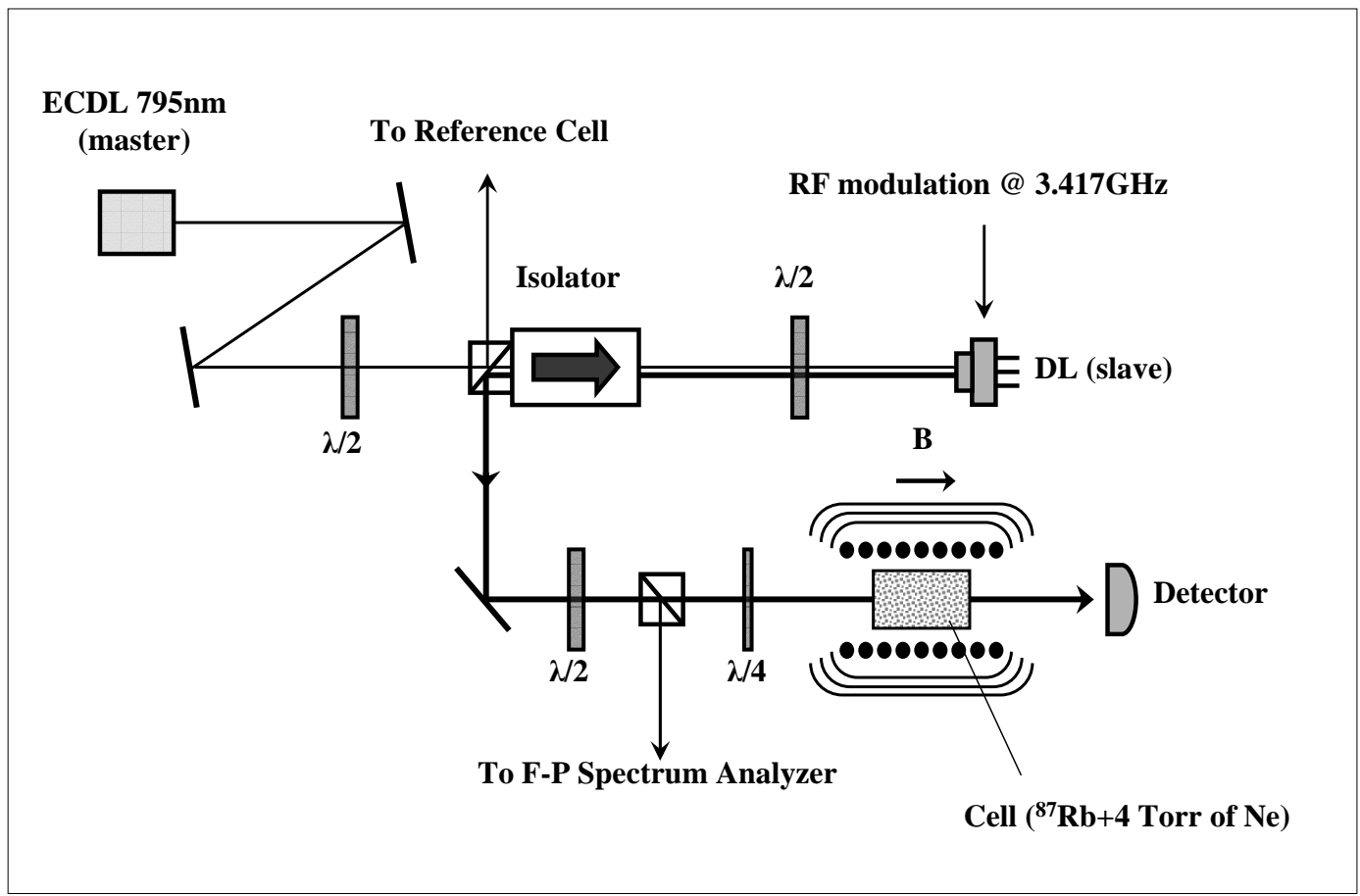}}}
\caption{Experimental setup. Cell contains $^{87}$Rb and 4 Torr of Ne buffer gas.} \label{fig2}
\end{figure*}

\begin{figure*}[h]\centerline{\scalebox{1.5}{\includegraphics{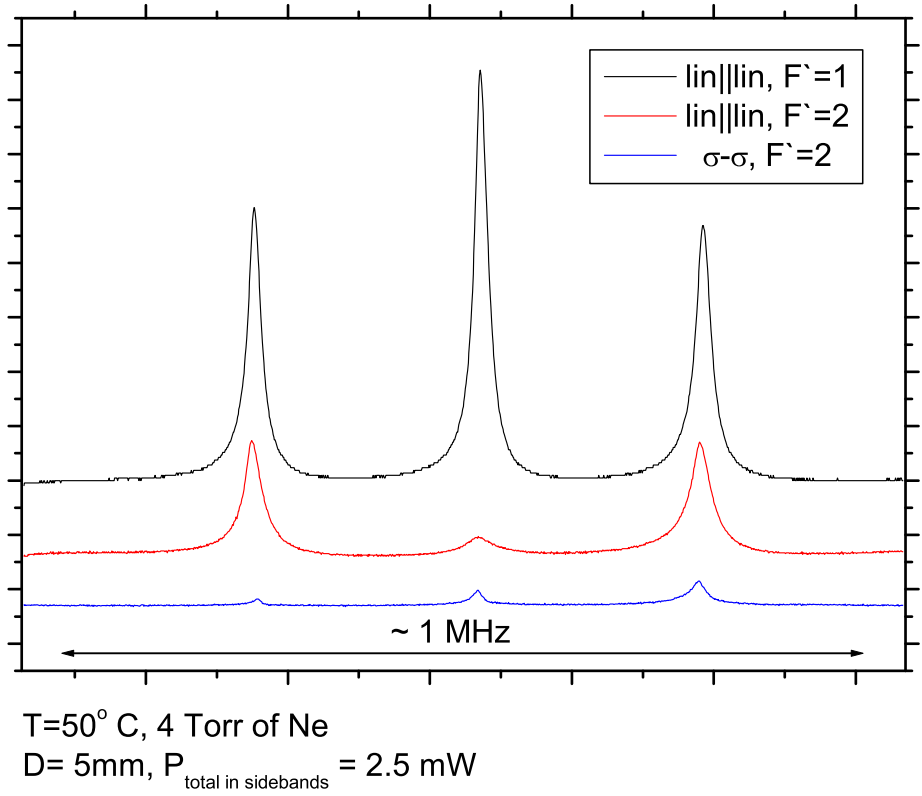}}}
\caption{CPT resonances in case of different excitation schemes.} \label{fig3}
\end{figure*}

\begin{figure*}[h]\centerline{\scalebox{1.}{\includegraphics{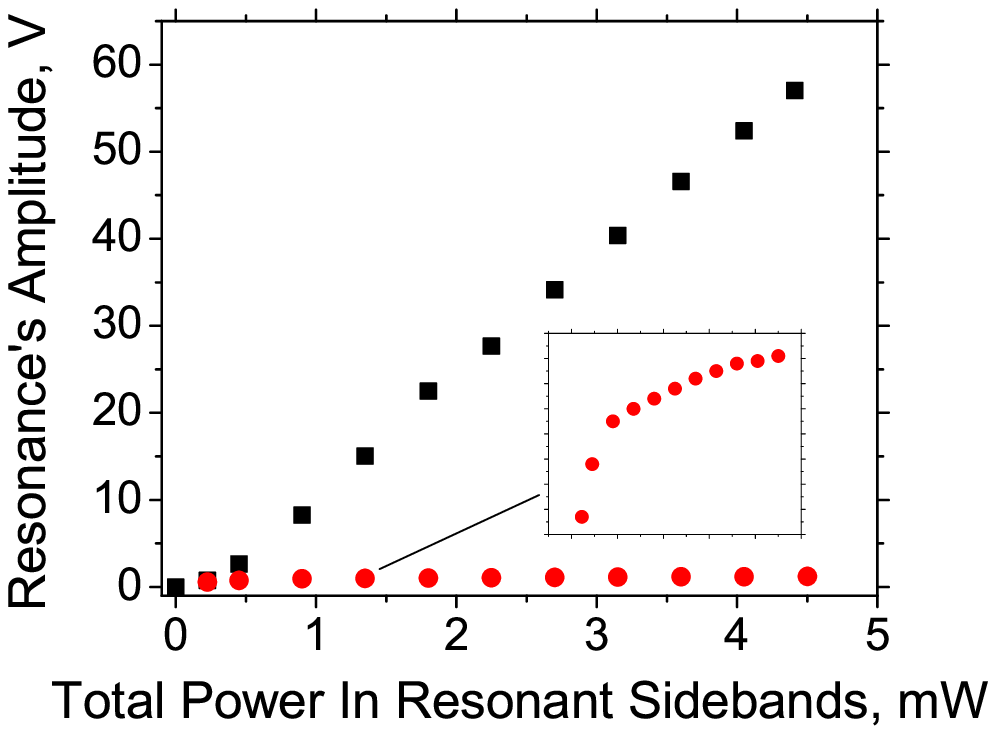}}}
\caption{Amplitude of the CPT resonance in case of {\em lin}$||${\em lin} (black squares) and $\sigma$-$\sigma$
(red circulars) excitation schemes.} \label{fig4}
\end{figure*}

\begin{figure*}[h]\centerline{\scalebox{1.}{\includegraphics{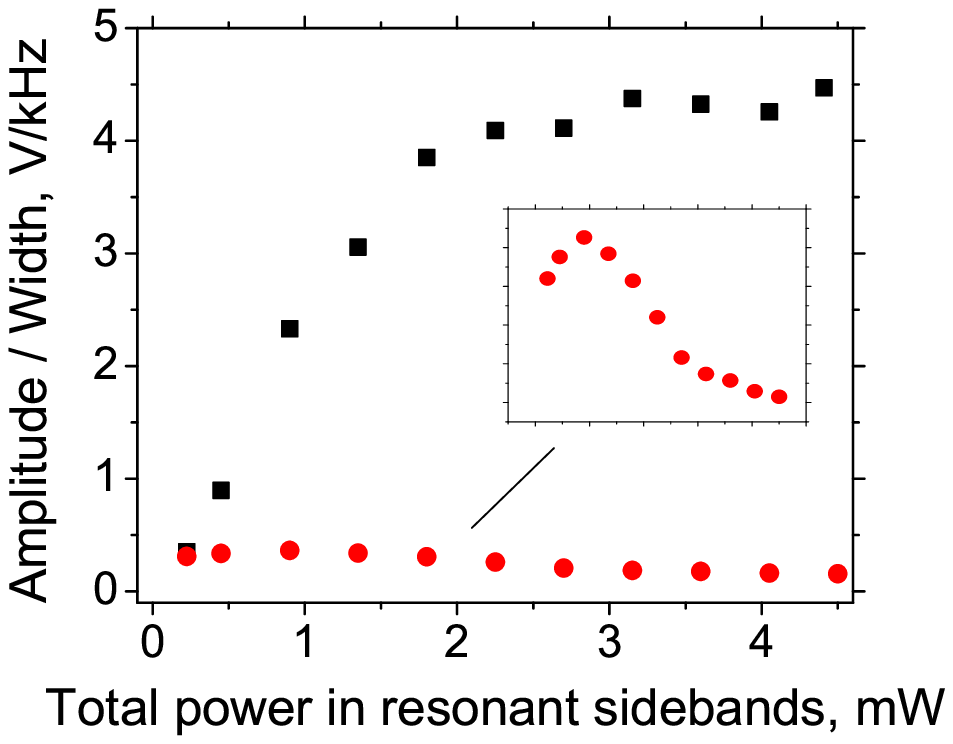}}}
\caption{Amplitude-to-width ratio of the CPT resonance in case of {\em lin}$||${\em lin} (black squares) and
$\sigma$-$\sigma$ (red circulars) excitation schemes.} \label{fig5}
\end{figure*}

\begin{figure*}[h]\centerline{\scalebox{1.}{\includegraphics{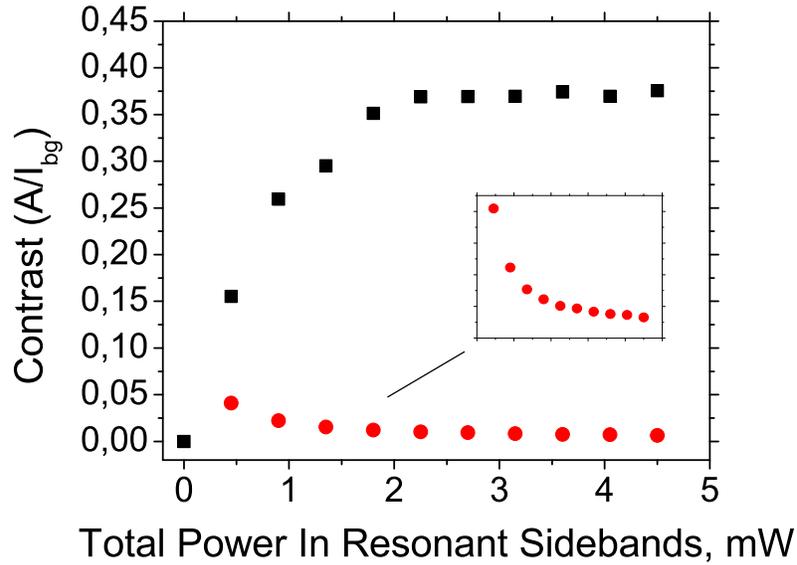}}}
\caption{Contrast of the CPT resonance in case of {\em lin}$||${\em lin} (black squares) and $\sigma$-$\sigma$
(red circulars) excitation schemes.} \label{fig6}
\end{figure*}

\begin{figure*}[h]\centerline{\scalebox{2.}{\includegraphics{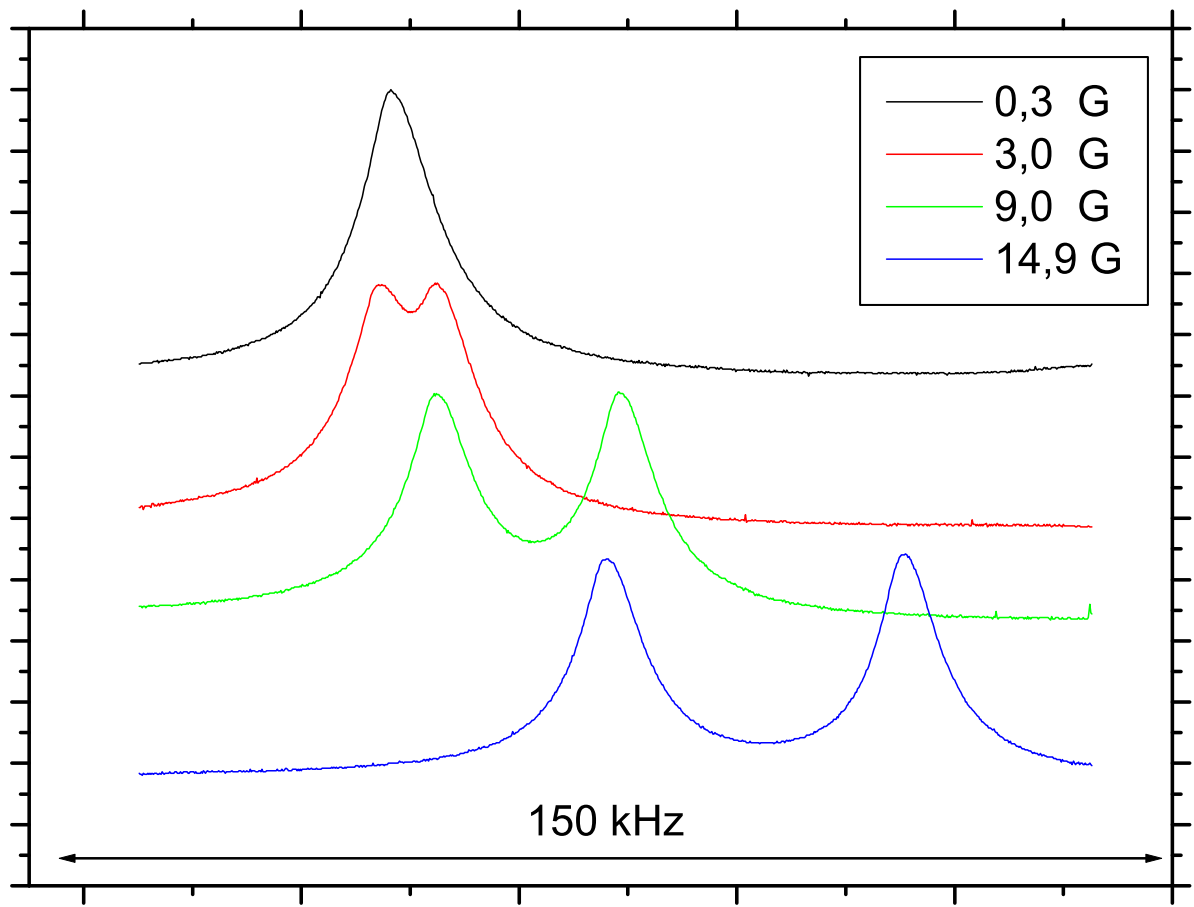}}}
\caption{CPT resonances for {\em lin}$||${\em lin} excitation through $F_e$=1 at different values of applied
longitudinal magnetic field.} \label{fig7}
\end{figure*}

\end{document}